\documentclass[12pt]{article}
\input{tcilatex}
\begin{document}

\begin{titlepage}
\begin{center}
%\hfill  \quad DFF 318/8/98 \\
\hfill  \quad hep-th/9902026 \\
\vskip .5 in
{\Large $\widehat{U(1)}\times\widehat{SU(m)}_{1}$ Theory and $c=m$
$W_{1+\infty}$ Minimal}
\vskip 0.15in
{\Large  Models in the Hierarchical Quantum Hall Effect}

\vskip 0.4in
Marina ~HUERTA \\%[.2in]
{\em Centro At\'omico Bariloche and Instituto Balseiro,
C. N. E. A. and Universidad Nacional de Cuyo,
8400 - San Carlos de Bariloche,
R\'{\i}o Negro, Argentina}
\end{center}
\vskip .1 in
\begin{abstract}
Two classes of Conformal Field Theories have been proposed to
describe the Hierarchical Quantum Hall Effect:
the multi-component bosonic theory, characterized by the symmetry
$\widehat{U(1)}\times\widehat{SU(m)}_{1}$ and the
$W_{1+\infty}$ minimal models with central charge $c=m$. In spite of having
the same spectrum of edge excitations,
they manifest differences in the degeneracy
of the states and in the quantum statistics, which
call for a more detailed comparison between them. Here, we describe
their detailed relation for the general case, $c=m$ and extend the
methods previously published for $c \leq 3$. Specifically, we obtain
the reduction in the number
of degrees of freedom from the multi-component Abelian theory to
the minimal models by decomposing
the characters of the $\widehat{U(1)}\times\widehat{SU(m)}_{1}$
representations into those of the $c=m$ ${W_{1+\infty}}$ minimal models. Furthermore, we find
the Hamiltonian whose renormalization group flow interpolates
between the two models, having the $W_{1+\infty}$ minimal models
as infra-red fixed point.

\end{abstract}
\vskip 1.cm
\vfill
\hfill January 1999
\end{titlepage}

\section{Introduction}

\strut

The low energy edge excitations of incompressible quantum Hall fluids \cite
{prange} are one-dimensional chiral waves which propagate on the boundary of
the sample. These could be described by $(1+1)$ dimensional effective
conformal field theories \cite{wen}. Moreover, this effective approach to
describe the low energy physics, is in accordance with the high precision
and universality found experimentally in the structure of the plateaus for
the Hall conductivity.

The filling fractions of the more stable plateaus are given by the Jain
series \cite{jain}:

\begin{equation}
\nu =\frac{m}{mp+1}\;\text{,}  \label{Jain}
\end{equation}
with $m=1,2,3,...$ and $p=2,4,6,...$.This corresponds to the majority of the
filling fractions observed experimentally in the range $0<\nu <1$ ;
therefore, its study and comprehension is very relevant in the Quantum Hall
Effect problem. For the Laughlin case with filling fraction $\nu =\frac{1}{%
p+1}$ ($m=1$), the corresponding conformal theory is the Abelian chiral
boson one, which describes the universal properties of the quasi-particle
excitations; these are in agreement with the Laughlin's microscopic theory
for the charge spectrum $Q$ and the quantum statistics $\theta /\pi $ \cite
{laugh}.

In contrast to the Laughlin case, general hierarchical Hall plateaus, whose
filling fractions are given by the Jain formula (\ref{Jain}) for general $%
m\neq 1$, are not well understood yet. At the moment, two classes of
conformal field theories have been proposed. The first is the
multi-component bosonic theory, characterized by the Abelian current algebra 
$\widehat{U(1)}^{m}$ (more specifically $\widehat{U(1)}\times \widehat{SU(m)}%
_{1}$ ), which is a natural generalization of the Abelian case made by $m$
copies of the one component theory \cite{abe}. The second one is given by
the minimal models of the $W_{1+\infty }$ symmetry algebra \cite{ctz5},
based on the dynamical symmetry of classical incompressible fluids under
area-preserving diffeomorphisms \cite{ctz}.

The spectra of these two types of theories are identical, but they differ in
the degeneracy of their excitations. This can be tested by computing
numerically the energy spectrum for the Hall states of a few electron
system. This kind of study has shown a reduction in the number of degrees of
freedom compared to the Abelian theory, which is in agreement with the
prediction of the $W_{1+\infty }$ minimal models \cite{cmsz}. This result
motivates our analysis of the general hierarchical problem.

Further important differences between the two theories are that the minimal
models exist for the hierarchical plateaus only, and that they possess a
single Abelian charge, rather than the $m$ charges of the Abelian
multi-component theory. Moreover, the minimal model quasi-particles possess
a type of non-Abelian statistics: actually, they are labelled by the weights
of the $SU(m)$ Lie algebra instead; the multi-component theories, on the
contrary, they are characterized by the $(m-1)$ Abelian charges.

The connection between the two types of theories has been explicitly found
in Ref.\cite{zcham} for the $c=2$ and $c=3$ cases. By introducing the
bosonic Fock space and by using the algebraic methods of representation
theory, it is possible to show that the minimal models correspond to a
sub-set of the states of the Abelian theory. This reduction in the number of
degrees of freedom can be performed by adding an interaction term to the
original Hamiltonian: this is relevant in the renormalization-group sense,
such that the corresponding flow interpolates from the multi-component
abelian to the minimal theories in the infra-red.

The purpose of this paper is to generalize these results to any $c=m>3$.

\smallskip

\strut

In section 2 we include a brief summary of the $c=2$ theory results. In
section 3 we present the decomposition of the $\widehat{U(1)}\times \widehat{%
SU(m)}_{1}$ characters into those of the $c=m$ $W_{1+\infty }$ minimal
models and the Hamiltonian that interpolates between the two theories..

\section{The $c=2$ case: Minimal model and two component Abelian theory}

\strut

\strut The two component Abelian theories are built out of two chiral bosons 
$\varphi ^{(i)}$, $i=1,2$, which are defined on the edge of the Hall sample,
which we take as a disk parametrized by a radius $R$ and an angle $\theta $ 
\cite{wen}. The chiral currents $J^{(1)}$ and $J^{(2)}$ associated with
these bosonic fields describe the edge excitations; their Fourier modes $%
\alpha _{n}^{(i)}$, $i=1,2$, satisfy the Abelian $\widehat{U(1)}\times 
\widehat{U(1)}$ current algebra

\begin{equation}
\left[ \alpha _{n}^{(i)},\alpha _{n}^{(j)}\right] =\delta ^{ij}n\text{ }%
\delta _{n+m,0}\;\text{.}
\end{equation}
The associated generators of the Virasoro algebra satisfy

\begin{equation}
\left[ L_{n}^{(i)},L_{n}^{(j)}\right] =\delta ^{ij}\left\{
(n-m)L_{n+m}^{(i)}+\frac{c}{12}n(n^{2}-1)\delta _{n+m,0}\right\} \;\text{,}
\end{equation}
with $c=1$, where $L_{n}^{(i)}$ are obtained from the Sugawara construction

\begin{equation}
L_{n}^{(i)}=\frac{1}{2}\alpha _{0}^{(i)}+\sum\limits_{n=0}^{\infty }\alpha
_{-n}^{(i)}\alpha _{n}^{(i)}\;\text{.}
\end{equation}

The eigenvalues of $\alpha _{0}^{(i)}$ and $L_{0}^{(i)}$ characterize the
highest weight representation of the Abelian algebra. Here, it is convenient
to define $\alpha _{n}=\alpha _{n}^{(1)}+\alpha _{n}^{(2)}$ and $%
L_{n}=L_{n}^{(1)}+L_{n}^{(2)}$ with $c=2$ and the same algebra. The
fractional statistics $\theta /\pi $ and the charge of the highest weight
representations are, therefore, given by the eigenvalues of $%
(L_{0}^{(1)}+L_{0}^{(2)})$ and $(\alpha _{0}^{(1)}+\alpha _{0}^{(2)})$,
respectively. It is possible to extend the $\widehat{U(1)}\times \widehat{%
U(1)}$ symmetry to $\widehat{U(1)}_{diagonal}\times \widehat{SU(2)}_{1}$
choosing an appropriate set of representations closed under the fusion
rules. This choice reproduces the Jain values of the filling fraction $\nu
=2/5,$ $2/3,...$. More precisely, the $\widehat{U(1)}\times \widehat{U(1)}$
representations are characterized by the eigenvalues of the $\alpha
_{0}^{(i)}$, zero modes of the chiral currents, which span a two-dimensional
lattice $\Gamma $ ; each lattice represents a single theory. We can choose a
particular class of lattices whose filling fractions match the Jain series
and whose representations allow the previously mentioned extension of the
symmetry $\widehat{U(1)}\times \widehat{U(1)}$ $\rightarrow $ $\widehat{U(1)}%
_{diagonal}\times \widehat{SU(2)}_{1}$ \cite{abe}.

The corresponding spectrum of quasi-particles with $\nu =2/(2p+1)$ is found
to be \cite{abe}:

\begin{eqnarray}
Q &=&\frac{n_{1}+n_{2}}{2p+1}\text{\hspace{0.15in}\ \ \ \ \ \ \ \ \ \ \ \ \
\ \ \ \ \ \ }n_{1},n_{2}\in \mathbf{Z}\text{ and }p=2,4,...\text{\ ,} 
\nonumber \\
h &=&\frac{1}{2}\frac{\theta }{\pi }=\frac{1}{2}(n_{1}^{2}+n_{2}^{2})-\frac{1%
}{2(2p+1)}(n_{1}+n_{2})^{2}\text{\hspace{0.15in}\ \ \ \ }n_{1},n_{2}\in 
\mathbf{Z\;}\text{\textbf{.}}  \label{esp1}
\end{eqnarray}

\strut With a simple change of variables, this spectrum can be decomposed
into charged and neutral sectors. The new integer quantum numbers $l$ and $n$
are defined by $n_{1}+n_{2}=2l+\alpha $ and $n_{1}-n_{2}=2n+\alpha $ , with $%
\alpha =0,1$; the spectrum of charges and quantum statistics is divided into
two parts with ``parity '' $\alpha $ \cite{zcham}

\emph{-Type I : }$\alpha =0$

\begin{eqnarray}
Q &=&\frac{2l}{2p+1}\text{\hspace{0.2in}\ \ \ \ \ \ \ \ \ \ \ \ \ \ \ \ }%
l\in \mathbf{Z}\text{ and }p=2,4,..\;\text{,}  \nonumber \\
h &=&\frac{1}{2}\frac{\theta }{\pi }=\frac{1}{2p+1}l^{2}+n^{2}\text{\hspace
{0.2in} \ \ \ \ \ \ \ }l,n\in \mathbf{Z\;}\text{\textbf{.}}
\end{eqnarray}
$.$

\emph{-Type II : }$\alpha =1$

\begin{eqnarray}
Q &=&\frac{2}{2p+1}(l+\frac{1}{2})\hspace{0.2in}\;\;\;\;\;\;\;\;\;\;\;\;\;\
;\;l\in \mathbf{Z}\text{ and }p=2,4,...\;\text{,}  \nonumber \\
h &=&\frac{1}{2}\frac{\theta }{\pi }=\frac{1}{2p+1}\left( l+\frac{1}{2}%
\right) ^{2}+\frac{\left( 2n+1\right) ^{2}}{4}\text{\hspace{0.2in}\ \ \ \ \
\ }l,n\in \mathbf{Z\;}\text{\textbf{.}}  \label{esp2}
\end{eqnarray}

After this change of variables, the neutral and charged excitations are
indeed factorized.

The number $l$ counts the units of fractional charge and the number $n$
labels the neutral quasi-particles. In this context, $p$ has a clear meaning
related to the compactification radius of the field $(\varphi ^{(1)}+\varphi
^{(2)})$, which represents the charged sector of the theory ($\widehat{U(1)}%
_{diagonal}$).

It is useful to introduce the currents and the Virasoro generators whose
eigenvalues are labelled by the new quantum numbers $l$ and $m$:

\begin{equation}
\begin{array}{llcl}
J=J^{(1)}+J^{(2)}\ , & {\frac{J_{0}}{\sqrt{2p+1}}} & \longrightarrow & Q={%
\frac{2\ell }{2p+1}}\ ; \\ 
J^{3}={\frac{1}{2}}\left( J^{(1)}-J^{(2)}\right) \ , & J_{0}^{3} & 
\longrightarrow & n\ ; \\ 
L=L^{(1)}+L^{(2)}=L^{Q}+L^{S}\ , & L_{0} & \longrightarrow & {\frac{\ell ^{2}%
}{2p+1}}+n^{2}\ ; \\ 
L^{Q}={\frac{1}{4}}\ :\ \left( J\right) ^{2}\ :\ ,\quad L^{S}=\ :\ \left(
J^{3}\right) ^{2}\ :\ . &  &  & 
\end{array}
\end{equation}

$\strut $

Each value of the spectrum corresponds to a highest weight state of the
Abelian current algebra $\widehat{U(1)}\times $ $\widehat{U(1)}$; each
highest weight state has an infinite tower of states on top generated by the
action of $\alpha _{n}^{(i)},n<0,i=1,2.$

Up to now we did not make explicit the extended symmetry $\widehat{U(1)}%
\times \widehat{SU(m)}_{1}$. The natural way to proceed is to incorporate
two additional chiral currents of dimension one \cite{cft}:

\begin{equation}
J^{\pm }=\ :\exp \left( \pm i\sqrt{2}\varphi \right) :\;\text{,}\ 
\end{equation}
with $\varphi =\frac{1}{2}\left( \varphi ^{(1)}-\varphi ^{(2)}\right) $.
Computing the commutators of the Fourier modes of $J^{\pm },J^{3}$ , we get

\begin{eqnarray}
\left[ \ J_{n}^{a}\ ,\ J_{m}^{b}\ \right] &=&i\epsilon ^{abc}J_{n+m}^{c}+%
\frac{k}{2}\delta ^{ab}\delta _{n+m,0}\ ,\quad k=1,\ a,b,c=1,2,3\text{,} 
\nonumber \\
\left[ \ L_{n}^{S}\ ,\ J_{m}^{a}\ \right] &=&-m\ J_{n+m}^{a}\;\text{,}
\end{eqnarray}
which is the $SU(2)_{1}$current algebra. Therefore, it is clear that the
extended symmetry is a consequence of the choice of the particular lattice $%
\Gamma $ .

So far, we described the Hall fluid with $\nu =\frac{2}{2p+1}$ in terms of
the rational conformal field theory $\widehat{U(1)}_{diagonal}\times 
\widehat{SU(2)}_{1}$ . Next, we compare this theory with the alternative
minimal models of the $W_{1+\infty }$ algebra \cite{zcham}. A detailed
description of the incompressible fluids by $w_{\infty }$ transformations
and the definitions of the $W_{1+\infty }$ algebra is given in \cite{ctz5}.
Here, we just summarize the main properties which are useful in our analysis.

In general, the $c=m$ $W_{1+\infty }$ unitary representations are of two
types: \textit{generic or degenerate. }The generic $W_{1+\infty }$ theories
are equivalent to chiral-boson theories; these have the same spectra of
filling fractions, fractional charge and statistic of excitations. For the $%
c=2$ case the representations of the Abelian algebra $\widehat{U(1)}\times 
\widehat{U(1)}$ with charges $(n^{1},n^{2})$ (eigenvalues of $J_{0}^{1}$ and 
$J_{0}^{2}$ ) are one-to-one equivalent to those of the $W_{1+\infty }$
algebra satisfying $(n^{1}-n^{2})\notin Z$ . On the other hand, the
degenerate $W_{1+\infty }$ representations are not equivalent to the Abelian
ones, and are contained into them, their charges $(n^{1},n^{2})$ satisfy $%
(n^{1}-n^{2})\in Z$ . A remarkable property of the degenerate
representations is that they correspond one-to-one to the hierarchical
plateaus and are thus physically relevant \cite{ctz5}. The $c=m$ $%
W_{1+\infty }$ degenerate representations are equivalent to the
representations of the $\widehat{U(1)}\times \mathcal{W}_{m}$ algebra, where 
$\mathcal{W}_{m}$ is the Zamolodchikov-Fateev-Lukyanov algebra at $c=m-1$ 
\cite{fateev}. The $SU(m)$ quantum number associated with the neutral sector
we mentioned before arises from the fusion rules of the $\mathcal{W}_{m}$
algebra, which are isomorphic to the tensor product of the $SU(m)$ Lie
algebra representations.

In the $c=2$ case \cite{zcham}, we have

\begin{equation}
\mathrm{degenerate}\ W_{1+\infty }\ \ \mathrm{reps}\ =\ \widehat{U(1)}\times 
\mathrm{Vir}\ \ \mathrm{reps}\ 
\end{equation}
where the $\widehat{U(1)}$ Abelian algebra corresponds to the charged sector
and the $\mathcal{W}_{2}\approx $ Virasoro algebra accounts for the neutral
one.

The spectrum of the $W_{1+\infty }$ minimal models is the same as that of
the $\widehat{U(1)}_{diagonal}\times \widehat{SU(2)}_{1}$ theories (\ref
{esp1}-\ref{esp2}), with the integer $n$ restricted to $n=0,1,2,..$.: each
point of the lattice defined by the numbers $n,l$, corresponds to one
degenerate $W_{1+\infty }$ representation, while in the $\widehat{U(1)}%
\times \widehat{SU(2)}_{1}$ this is a $\widehat{U(1)}\times \widehat{U(1)}$
representation with the same Virasoro weight.

Using the characters of the $\widehat{SU(2)}_{1}$representations with
isospin $\sigma =\alpha /2=0,1/2$ and the Virasoro representations $%
h=n^{2}/4 $ \cite{itz} \cite{cft} it is possible to count the number of edge
excitations in the neutral sector, which is different in the two theories 
\cite{zcham}. Their expressions are:

\begin{eqnarray}
\chi _{\sigma =0}^{\widehat{SU(2)_{1}}} &=&{\frac{1}{\eta (q)^{3}}}\
\sum_{k\in \mathbf{Z}}\left( 6k+1\right) \ q^{(6k+1)^{2}/12}\;\text{,} 
\nonumber \\
\chi _{\sigma =1/2}^{\widehat{SU(2)_{1}}} &=&{\frac{1}{\eta (q)^{3}}}\
\sum_{k\in \mathbf{Z}}\ \left( 6k+2\right) \ q^{(6k+2)^{2}/12}\;\text{,} 
\nonumber \\
\chi _{L_{0}^{S}=n^{2}/4}^{\mathrm{Vir}} &=&{\frac{q^{n^{2}/4}\left(
1-q^{n+1}\right) }{\eta (q)}}={\frac{q^{n^{2}/4}-q^{(n+2)^{2}/4}} {\eta (q)}%
\;.}
\end{eqnarray}

\strut It is convenient to relate them to the characters of $\widehat{U(1)}$
representations

\begin{equation}
\chi _{L_{0}=h}^{\widehat{U(1)}}={\frac{q^{h}}{\eta (q)}}\ ,\qquad \eta
(q)=q^{1/24}\ \prod_{k=1}^{\infty }\left( 1-q^{k}\right) \;\text{,}
\end{equation}

as follows:

\begin{eqnarray}
\chi _{\sigma =0}^{\widehat{SU(2)_{1}}}\ &=&\ {\frac{1}{\eta (q)}}\sum_{k\in 
\mathbf{Z}}\ q^{k^{2}}\text{, \ \ \ \ }\chi _{\sigma =1/2}^{\widehat{%
SU(2)_{1}}}\ =\ {\frac{1}{\eta (q)}}\sum_{k\in \mathbf{Z}}\ q^{(2k+1)^{2}/4}%
\text{,} \\
\text{\ \ \ \ \ }\chi _{J_{0}^{3}=n/2}^{\widehat{U(1)}} &=&\sum_{\ell
=0}^{\infty }\ \chi _{L_{0}^{S}=(n+2\ell )^{2}/4}^{\mathrm{Vir}}\;\text{.} 
\nonumber
\end{eqnarray}

$\ $

From these formulae, the characters of the two theories are related as
follows \cite{ctz}:

\begin{eqnarray}
\chi _{\sigma =0}^{\widehat{SU(2)_{1}}} &=&\sum_{k=0}^{\infty }(2k+1)\ \chi
_{L_{0}^{S}=(2k)^{2}/4}^{\mathrm{Vir}}\;\text{,} \\
\text{\ \ \ \ \ }\chi _{\sigma =1/2}^{\widehat{SU(2)_{1}}}
&=&\sum_{k=0}^{\infty }(2k+2)\ \chi _{L_{0}^{S}=(2k+1)^{2}/4}^{\mathrm{Vir}%
}\;\text{.}  \nonumber
\end{eqnarray}

The above relations correspond to the following decomposition of $\widehat{%
SU(2)_{1}}$ representations into $\widehat{U(1)}$ and Virasoro ones:

\begin{equation}
\{\sigma =0\}_{\widehat{SU(2)_{1}}}=\sum_{k\in \mathbf{Z}\ \mathrm{even}}\
\left\{ J_{0}^{3}={\frac{k}{2}}\right\} _{\widehat{U(1)}}\ =\
\sum_{s=0}^{\infty }\ \left( 2s+1\right) \left\{ L_{0}^{S}=s^{2}\right\} _ {%
\mathrm{Vir}}\;\text{,}
\end{equation}

\strut 
\[
\left\{ \sigma ={\frac{1}{2}}\right\} _{\widehat{SU(2)_{1}}}=\sum_{k\in 
\mathbf{Z}\ \mathrm{odd}}\ \left\{ J_{0}^{3}={\frac{k}{2}}\right\} _ {%
\widehat{U(1)}}=\sum_{s=1/2,\ s\in \mathbf{Z}^{+}+1/2}^{\infty }\ \left(
2s+1\right) \left\{ L_{0}^{S}=s^{2}\right\} _{\mathrm{Vir}}\;\text{.} 
\]

These expressions show: first, that the two $\widehat{SU(2)_{1}}$
representations sum up the neutral spectrum of the Abelian theory in the two
sectors, type I and II, for $\sigma =0$ and $1/2$ respectively; second, that
the multiplicities of the decomposition into Virasoro representations match
the familiar multiplicities $(2s+1)$ of the $SU(2)$ Lie algebra
representations. We see that the minimal models contains the Virasoro $%
h=s^{2}$ representations with multiplicity 1, while the $SU(2)_{1}$ theory
contains $(2s+1)$ of them. Therefore, one can obtain the minimal models from
the $\widehat{U(1)}\times \widehat{SU(m)}_{1}$ theory by selecting one state
for each $SU(2)$ multiplet: this can be done by imposing the constraint on
the states \cite{zcham}: 
\begin{equation}
J_{0}^{-}\ |\ \mathrm{minimal\ state}\ \rangle \ =\ 0\ \;\text{,}
\label{cons}
\end{equation}
The $\widehat{U(1)}\times \widehat{SU(m)}_{1}$ states which satisfy the
constraint belong to the minimal models, and the others are projected out.
Therefore, the $W_{1+\infty }$ minimal models are defined by the $\widehat{%
U(1)}\times \widehat{SU(m)}_{1}$ Hilbert space plus the constraint (\ref
{cons}). The evident consequence of the incorporation of the constraint is
the breaking of the $SU(2)$ symmetry and the fixing of the $U(1)$ quantum
number $m=-s$ .

The Hamiltonian approach of Ref \cite{zcham} is an equivalent way to present
the results we just described. First, let us consider the standard $\widehat{%
U(1)}\times $ $\widehat{U(1)}$ Hamiltonian

\begin{equation}
H=\frac{1}{R}\left( \ v\ L_{0}^{Q}\ +\ v^{\prime }\ L_{0}^{S}\ -\ \frac{1}{12%
}\ \right) \text{,}
\end{equation}
$\ \ $ Next, we add a (non-local) relevant term:

\begin{equation}
\Delta H=\ \gamma \ J_{0}^{+}\ J_{0}^{-}\ \qquad \qquad \gamma \in [0,\infty
)\;\text{,}
\end{equation}
This is diagonal in the previous $(m,s^{2})$ basis, and assigns the energy $%
\ \gamma [m(m+1)-s(s+1)]$ to each state of the $SU(2)$ multiplets. The
coupling constant $\gamma $ has dimension of a mass which assures the
relevancy of the term in the renormalization-group sense. For $\gamma
\rightarrow \infty $ , it selects the lowest weight in each $SU(2)$
multiplet, $m=-s$, in other words, implements the constraint \cite{zcham}.

Therefore, the infra-red limit of the theory defined by the Hamiltonian $%
H+\Delta H$ is the $c=2$ $W_{1+\infty }$ minimal model. This infra-red fixed
point is known exactly and is conformally invariant because $\left[
J_{0}^{+}\ J_{0}^{-}\ ,L_{0}^{S}\right] =0$ .The trajectory in the theory
space defined by the action of the renormalization-group interpolates
between the $\widehat{U(1)}\times \widehat{SU(2)}_{1}$theory $(\gamma
\rightarrow 0)$ and the $c=2$ minimal model $(\gamma \rightarrow \infty )$ .
The fact that the conformal field theories are describing the low energy
edge excitations, makes physically significant this infra-red limit, which
is reached without any fine-tuning.

\section{$c=m$ W$_{1+\infty }$ Minimal Models and $\widehat{U(1)}\times 
\widehat{SU(m)}_{1}$ Theories\strut}

The purpose of this paper is to generalize the previous analysis to describe
the $c=m$ case. In particular, we will study the reduction in the number of
the degrees of freedom from the $\widehat{U(1)}\times \widehat{SU(m)}_{1}$
to the $c=m$ $W_{1+\infty }$ minimal models. We start with the algebraic
analysis of the corresponding characters, and we complete the description
including the Hamiltonian point of view.

As we mentioned in the Introduction, the $\widehat{U(1)}\times \widehat{SU(m)%
}_{1}$ theory can be built out of $m$ chiral bosons defined on the edge of
the Hall sample. The zero modes of the chiral currents $J^{(i)},$ $i=1,...m$%
, associated with these bosons, span a $m$-dimensional lattice. From the set
of possible lattices, we choose the particular class, in which the filling
fractions correspond to the Jain series $\nu =m/(mp+1)$. These lattices
imply the symmetry extension from $\widehat{U(1)}^{m}\rightarrow \widehat{%
U(1)}\times \widehat{SU(m)}_{1}$.

The spectrum of the theory is given by \cite{abe}:

\begin{eqnarray}
Q &=&\frac{1}{pm+1}\sum\limits_{i=1}^{m}n_{i}\;\text{,}  \nonumber \\
2h &=&\frac{\theta }{\pi }=\sum\limits_{i=1}^{m}n_{i}^{2}-\frac{p}{mp+1}%
(\sum\limits_{i=1}^{m}n_{i})^{2}\hspace{0.2in}\;\;\;\;\;\;\;n_{i}\in \mathbf{%
Z\;}\text{\textbf{.}}  \label{spect}
\end{eqnarray}
In the lattice $\Gamma $ the norm of the vectors is $h$, the total Virasoro
dimension, and each point identifies a highest weight representation of $%
\widehat{U(1)}^{m}$. In order to perform the factorization into charged and
neutral sectors, we introduce a set of definitions related with the $SU(m)$
Lie Algebra. Consider the $(m-1)$-dimensional sub-lattice $P$ generated by
the fundamental weight $\mathbf{\Lambda }^{i}$ of $SU(m)$ and $\mathbf{%
\Lambda }=\sum\limits_{i=1}^{m-1}l_{i}\mathbf{\Lambda }^{i}\in P$ a vector
of $P$. A standard basis for the weights is \cite{wyb}:

$\mathbf{u}^{1}=\left( \;\frac{1}{\sqrt{2}},\frac{1}{\sqrt{6}},\frac{1} {%
\sqrt{12}},........,\;\frac{1}{\sqrt{k(k-1)}},.....,\frac{1} {\sqrt{m(m-1)}}%
\right) $

$\mathbf{u}^{2}=\left( -\frac{1}{\sqrt{2}},\frac{1}{\sqrt{6}}%
,...................................,\frac{1}{\sqrt{m(m-1)}}\right) $

$.$

$.$

$\mathbf{u}^{k}=\left( \;0,.....,0,.............,-\frac{k-1}{\sqrt{k(k-1)} }%
,....,\frac{1}{\sqrt{m(m-1)}}\right) $

$\mathbf{u}^{m}=\left( \;0,.....,0,...............................,0,\frac
{1}{\sqrt{m(m-1)}}\right) $

with

\begin{equation}
\mathbf{\Lambda }^{i}=\sum\limits_{j=1}^{i}\mathbf{u}^{j}
\end{equation}
The simple positive roots $\mathbf{\alpha }_{i}$ dual to the $\mathbf{%
\Lambda }^{i}$ are given by

\begin{equation}
\mathbf{\alpha }_{i}=\mathbf{u}^{i}-\mathbf{u}^{i+1}
\end{equation}
with

\begin{equation}
\mathbf{\Lambda }^{i}.\mathbf{\alpha }_{j}=\delta _{j}^{i}\text{ ,}%
\;\;\;\;\;\;\;\;\;\;\;\;\;\;\;\;\;\;\;\;\;\text{ }\mathbf{\Lambda }^{i}.%
\mathbf{\Lambda }^{j}=i-\frac{ij}{m}\hspace{0.2in}i\leq j\;\text{,}\;
\end{equation}
and a vector in the root lattice is of the form $\mathbf{\gamma }%
=\sum\limits_{j=1}^{i}k_{i}\mathbf{\alpha }_{i}$. The relation between the
root lattice and the weight lattice is given by the Cartan matrix:

\begin{equation}
\mathbf{\alpha }_{i}=2\mathbf{\Lambda }^{i}-\mathbf{\Lambda }^{i-1}-\mathbf{%
\Lambda }^{i+1}
\end{equation}
The representations of $SU(m)$ Lie algebra can be characterized by a
positive highest weight vector $\mathbf{\Lambda }\in P^{+}$, in other words,
the representations are labelled by $(l_{1},...,l_{m-1})$ with $l_{i}\geq 0$%
. In addition, their $m$-ality is given by:

\begin{equation}
\alpha =\sum\limits_{i=1}^{m-1}i\text{ }l_{i\text{ }}\;\;\;\text{mod}(m) \;%
\text{,}  \label{m-ality}
\end{equation}

$\alpha =0,1,...m-1\;.$

Next, it is useful to introduce the following property

\begin{equation}
\sum\limits_{i=1}^{m-1}il_{i}=\sum%
\limits_{i=1}^{m-1}i(l_{i}+2k_{i}-k_{i-1}-k_{i+1})\;\ \ \ mod(m),  \label{a}
\end{equation}
which means that the $m$-ality is invariant under the addition of a vector
in the root lattice . We will denote $P_{\alpha }$ the $m$ sub-lattices of $%
P $ with the given value of $\alpha $ of the $m$-ality. Now, the
factorization into charged and neutral sectors is obtained by the change of
basis :

\begin{equation}
(n_{1},...n_{m})\rightarrow (l,l_{1},...,l_{m-1})\;\text{,}
\end{equation}
with $l=\sum\limits_{i=1}^{m}n_{i}$ and $l_{i}=n_{i}-n_{i-1}$. The spectrum (%
\ref{spect}) can be rewritten:

\begin{eqnarray}
Q &=&\frac{l}{pm+1}\;\text{,}  \nonumber \\
2h &=&\frac{\theta }{\pi }=h_{\mathbf{\Lambda }}+\frac{l^{2}}{2m(mp+1)}\ ;%
\text{,}
\end{eqnarray}

$\;$

\begin{equation}
h_{\mathbf{\Lambda }}=\frac{1}{2}\sum\limits_{i=1}^{m-1}(\frac{im-i^{2}}{m}%
)l_{i}^{2}+\sum\limits_{j=1}^{m-1}\sum\limits_{i=1}^{j-1}(i-\frac{ij}{m}%
)l_{i}l_{j}=\frac{1}{2}\left| \mathbf{\Lambda }\right| ^{2}\;\text{.}
\end{equation}
The integer $l$ identifies the charged sector and the $l_{i}$ the neutral
one.

For the purpose of comparison of the $\widehat{SU(m)}_{1}$ and $\mathcal{W}%
_{m}$ neutral spectrum, the last expression is not yet completely
appropriate; it is convenient to consider an $\alpha $-dependent change of
basis, making explicitly the $m$-ality of the theory. We modify the previous
transformation, (taking into account the property (\ref{a})) as follows:

For $\alpha =\alpha _{j}$, $j=0,1,...,m-1$:

\begin{equation}
ml+\alpha _{j}=\sum\limits_{i=1}^{m}n_{i}\;\text{,}  \label{li}
\end{equation}

\[
l_{i}=n_{i}-n_{i-1}=2k_{i}-k_{i-1}-k_{i+1}+l_{i}^{\prime },\;\;\;\;\;1\leq
i\leq m-1\ ;\text{,} 
\]
where $l_{i}^{\prime }=1$ if $i=j$ or $l_{i}^{\prime }=0$ if $i\neq j$.
Here, $\alpha _{i}=i$ with $i=0,...,m-1.$

The spectrum in the new basis is given by:

For $\alpha =\alpha _{j}$:

\begin{eqnarray}
Q_{\alpha _{j}} &=&\frac{ml+\alpha _{j}}{pm+1}\;\text{,}  \nonumber \\
2h_{\alpha _{j}} &=&\frac{\theta }{\pi }=h_{\mathbf{\Lambda }} +\frac{%
(ml+\alpha _{j})^{2}}{2m(mp+1)}\;\text{,}
\end{eqnarray}

with:

\begin{equation}
h_{\mathbf{\Lambda }}=\frac{1}{2}\sum\limits_{i=1}^{m-1}(\frac{im-i^{2}}{m}%
)l_{i}^{2}+\sum\limits_{j=1}^{m-1}\sum\limits_{i=1}^{j-1}(i-\frac{ij}{m}%
)l_{i}l_{j}\;\text{,}  \label{h}
\end{equation}

where $\mathbf{\Lambda }=(l_{1,}...,l_{m-1})$, with $l_{i}$ given by the
previous expression (\ref{li}).

We introduce the $SU(m)_{1}$ representations corresponding to $\mathbf{%
\Lambda }_{SU(m)_{1}}=(l_{1},l_{2},....,l_{m-1})$ with $\sum%
\limits_{i=1}^{m-1}l_{i}\leq 1$, and define $\mathbf{\Lambda }_{\alpha
_{i}}=(0,....1,....0)$, with the $i$-th component equal to 1 and $\mathbf{%
\Lambda }_{\alpha _{0}}=(0,....0,....0)$. This notation allows to rewrite
the previous change of basis in the following compact way:

For $\alpha =\alpha _{j}:$

\begin{eqnarray}
m\text{ }l+\alpha _{j} &=&\sum\limits_{i=1}^{m}\text{ }n_{i}\;\text{,} \\
\mathbf{\Lambda } &=&\mathbf{\Lambda }_{\alpha _{j}}+\mathbf{\gamma }\ ;%
\text{.}  \nonumber
\end{eqnarray}

Now, we are ready to compare the $\widehat{U(1)}\times \widehat{SU(m)_{1}}$
theory with the $c=m$ $W_{1+\infty }$ minimal models and study the reduction
in the number of the degrees of freedom. The characters of the
representations provide the relevant information as in the previous section,
because the projection from $\widehat{U(1)}\times $ $\widehat{SU(m)}_{1}$to
the minimal models can be obtained from the decomposition of the characters
of $\widehat{SU(m)}_{1}$. The $\widehat{SU(m)}_{1}$character for the $%
\mathbf{\Lambda }_{\alpha _{j}}$ representation is given by \cite{itz}:

\begin{equation}
\chi _{\mathbf{\Lambda }_{\alpha _{j}}}^{\widehat{SU(m)_{1}}}=\frac{1}{\eta
^{m^{2}-1}}\sum\limits_{k_{i}=-\infty }^{\infty }\dim (\mathbf{\Lambda }%
_{\alpha _{j}}+(m+1)\mathbf{\gamma })(q^{\frac{i}{2m-1}\left| 
\overrightarrow{p}_{\mathbf{\Lambda }_{\alpha _{j}}}+(m+1)\mathbf{\gamma }%
\right| ^{2}})\;\text{,}  \label{car1}
\end{equation}
with

$\overrightarrow{p}_{\mathbf{\Lambda }_{\alpha _{j}}}=\sum\limits_{i=1}^{m-1}%
\mathbf{\Lambda }_{i}+$ $\mathbf{\Lambda }_{\alpha _{j}}$; $\mathbf{\gamma }%
=\sum\limits_{i=1}^{m-1}k_{i}\alpha _{i}$, a vector of the root lattice and $%
\dim (\mathbf{\Lambda }_{\alpha _{i}}+(m+1)\mathbf{\gamma })$ the dimension
of the $SU(m)$ Lie Group representations given by:

\begin{eqnarray}
\dim (l_{1},...,l_{m-1}) &=&(1+l_{1}).............................(1+l_{m-1})
\nonumber \\
&&\times (1+\frac{l_{1}+l_{2}}{2})..........(1+\frac{l_{m-2}+l_{m-1}}{2}) 
\nonumber \\
&&\times ....\times (1+\frac{l_{1}+...+l_{m-1}}{m-1})\; \\
&=&\prod\limits_{j=1}^{m-1}\left( \prod\limits_{k=1}^{m-j}\left(
1+\sum\limits_{i=k}^{j+k+1}\frac{l_{i}}{j}\right) \right)
\end{eqnarray}

For the minimal models we consider the characters of the degenerate $%
\mathcal{W}_{m}$ representations \cite{kac}:

\begin{eqnarray}
\chi _{h_{\mathbf{\Lambda }}}^{\mathcal{W}_{m}} &=&\frac{q^{h_{\mathbf{%
\Lambda }}}}{\eta ^{m-1}}(1-q^{l_{1}+1})(1-q^{l_{2}+1})...(1-q^{l_{m-1}+1})
\label{car2} \\
&&\times (1-q^{l_{1}+l_{2}+2})(1-q^{l_{2}+l_{3}+2})...  \nonumber \\
&&\times ...  \nonumber \\
&&\times (1-q^{l_{1}+l_{2}+...+l_{m-1}+m-1})\;.  \nonumber
\end{eqnarray}
For both characters, (\ref{car1}) and (\ref{car2}), it is possible to
establish a relation with the characters of $\widehat{U(1)}^{m-1}$
representations, as follows:

\begin{equation}
\chi _{\mathbf{\Lambda }_{\alpha j}}^{SU(m)_{1}}=\frac{1}{\eta ^{m-1}}%
\sum\limits_{k_{i}\in \mathbf{Z}}q^{h_{\mathbf{\Lambda }}}\;\text{,}
\label{u1}
\end{equation}
where we consider the value of $h_{\mathbf{\Lambda }}$ (\ref{h}) which
corresponds to the sector $\alpha _{j}$ .

Finally, as in the $c=2$ case, we can relate the characters of the two
theories in a direct way:

\begin{equation}
\chi _{\mathbf{\Lambda }_{\alpha j}}^{\widehat{SU(m)_{1}}}=\sum%
\limits_{k_{i}\in \Phi }\dim (\mathbf{\Lambda }_{\alpha _{i}}+\mathbf{\gamma 
})\chi _{h_{\mathbf{\Lambda }}}^{\mathcal{W}_{m}}\text{.}
\end{equation}
with $\Phi =\{k_{i}/$ $2k_{i}-k_{i-1}-k_{i+1}+l_{i}^{\prime }\geq 0\}$ and
as in (\ref{u1}) $h_{\mathbf{\Lambda }}$ given in (\ref{h}) which
corresponds to the sector $\alpha _{j}$.

The two last identities have been checked with the help of the program
Mathematica \cite{math} for $m=3,4,5$ by an expansion in powers of $q$ . We
find for the general $c=m$ theories the following identities:

\begin{equation}
\{\Lambda _{\alpha _{j}}\}_{\widehat{SU(m)_{1}}}=\sum_{k_{i}\in \mathbf{Z}}\
\left\{ h=h_{\mathbf{\Lambda }}\right\} _{\widehat{U(1)^{m-1}}}\
=\sum_{k_{i}\in \Phi }\ \dim _{SU(m)}(\mathbf{\Lambda })\left\{ L_{0}=h_{%
\mathbf{\Lambda }}\right\} _{\mathcal{W}_{m}}
\end{equation}
We conclude again that the minimal models contains the $h_{\mathbf{\Lambda }%
} $ representations with multiplicity $1$, while the $SU(m)_{1}$contains $%
\dim _{SU(m)}(\mathbf{\Lambda })$ of them.

To complete our analysis, we consider the Hamiltonian of the $c=m$ $%
W_{1+\infty }$ minimal models, which is completely analogous to the $c=2$
case. The projection from the Abelian theory to the minimal models is
obtained by the application of the constraint on the states given by

\begin{equation}
E_{0}^{-\mathbf{\alpha }_{i}}\ |\ \mathrm{minimal\ state}\ \rangle \ =\ 0\ \
;\text{,}
\end{equation}
with $i=1,...,m-1$ and the $E_{0}^{-\mathbf{\alpha }_{i}}$ the ladder
operators of the $SU(m)$ Lie algebra. The analogous to $E_{0}^{-\mathbf{%
\alpha }_{i}}$ in the $SU(2)$ case is $J_{0}^{-}$, which select one state of
the multiplet -the lowest weight state. The relevant term we have to add to
the Hamiltonian to accomplish the projection has the form:

\begin{equation}
\Delta H=\gamma \sum\limits_{i=1}^{m-1}(E_{0}^{\mathbf{\alpha }_{i}}E_{0}^{-%
\mathbf{\alpha }_{i}})\;\text{,}
\end{equation}
because this is the unique term which annihilates the $SU(m)$ lowest-weight
state and is diagonal in the $\widehat{SU(m)}_{1}$ basis. As before, the
infra-red limit $\gamma \rightarrow \infty $ corresponds to the minimal
models, because other states in the hyper multiplets acquire an energy
proportional to $\gamma $. The Hamiltonian description, gives us a physical
framework in which it is possible to discuss the differences between the two
types of theories. The renormalization group action shows a natural link
between the theories relating the limits $\gamma \rightarrow 0$ and $\gamma
\rightarrow \infty $, corresponding to the Abelian theory and the minimal
models respectively.

\section{Acknowledgements}

\strut

I would like to thank Guillermo Zemba, for suggesting me this problem, for
discussions and for reading and commenting on the manuscript. I would also
like to thank Andrea Cappelli for useful comments. I also enjoyed
discussions with Horacio Casini.

This work was supported by a FOMEC fellowship of the Education Ministry of
Argentina.


\begin{thebibliography}{99}
\bibitem{prange}  For a review see: R. A. Prange and S. M. Girvin, \textit{%
The Quantum Hall Effect}, Springer Verlag, New York A.

\bibitem{wen}  For a review, see: X. G. Wen, \textit{Int. J. Mod. Phys.\ } 
\textbf{6 B} (1992) 1711; \textit{Adv. in Phys.} \textbf{\ 44} (1995) 405.

\bibitem{jain}  For a review see: J. K. Jain, \textit{Adv. in Phys.} \textbf{%
44} (1992) 105, \textit{Science} \textbf{266} (1994) 1199.

\bibitem{laugh}  R. B. Laughlin, \textit{Phys. Rev. Lett.\ } \textbf{50}
(1983) 1395; for a review see: R. B. Laughlin, \textit{Elementary Theory:
the Incompressible Quantum Fluid}, in \cite{prange}.

\bibitem{abe}  J. Fr\"{o}hlich and A. Zee, \textit{Nucl. Phys.\ } \textbf{%
364 B} (1991) 517; X.-G. Wen and A. Zee, \textit{Phys. Rev.\ } \textbf{46 B}
(1993) 2290. J. Fr\"{o}hlich and E. Thiran, \textit{J. Stat. Phys.} \textbf{%
76} (1994) 209; J. Fr\"{o}hlich, T. Kerler, U. M. Studer and E. Thiran, 
\textit{Nucl. Phys.\ } \textbf{B 453} (1995) 670.

\bibitem{ctz5}  A. Cappelli, C. A. Trugenberger and G. R. Zemba, \textit{%
Nucl. Phys.\ } \textbf{448} (1995) 470; for a review, see: \textit{Nucl.
Phys.\ } \textit{(Proc. Suppl.)} \textbf{B 45A} (1996) 112.

\bibitem{ctz}  A.Cappelli, C. A. Trugenberger and G. R. Zemba, \textit{Nucl.
Phys.\ } \textbf{396 B} (1993) 465; \textit{Phys. Lett.\ } \textbf{306 B}
(1993) 100; \textit{Phys. Rev. Lett.\ } \textbf{72} (1994) 1902; for a
review, see: A.Cappelli, G.V.Dunne, C.A.Trugenberger and G.R.Zemba, \textit{%
Nucl. Phys.\ } \textbf{B (Proc. Suppl.) 33 C} (1993) 21.(1990);

\bibitem{cmsz}  A. Cappelli, C. Mendez, J. M. Simonin and G. R. Zemba,
preprint cond-mat/9806238, \textit{Phys. Rev.\ } \textbf{B} in press.

\bibitem{zcham}  A. Cappelli and G.R. Zemba, hep-th 9808179, \textit{Nucl.
Phys}. \textbf{B}, in press.

\bibitem{cft}  P. Di Francesco, P. Mathieu and D. Senechal, \textit{%
Conformal Field Theories}, Springer-Verlag, (1996).

\bibitem{fateev}  V.A. Fateev and A.B. Zamolodchikov, \textit{Nucl. Phys}. 
\textbf{B 280} (1989) 644; V.A. Fateev and S.L. Lukyanov, \textit{Int. J.
Mod. Phys.} \textbf{A 3}(1988) 507.

\bibitem{itz}  C. Itzykson, \textit{Nucl. Phys.\ } \textit{(Proc. Suppl.)} 
\textbf{5 B} (1988) 150.

\bibitem{wyb}  B. G. Wybourne, \textit{Classical Groups for Physicists},
Wiley, New York (1974).

\bibitem{kac}  V. Kac and A. Radul, \textit{Comm. Math. Phys.} \textbf{157}
(1993) 429; E. Frenkel, V. Kac, A. Radul and W. Wang, \textit{Comm. Math.
Phys.\ } \textbf{170} (1995) 337; H. Awata, M. Fukuma, Y. Matsuo and S.
Odake, \textit{Prog. Theor. Phys. (Supp.)} \textbf{118} (1995) 343.

\bibitem{math}  S. Wolfram, \textit{Mathematica}, Addison-Wesley, New York
(1991)
\end{thebibliography}
\end{document}